\catcode`\@=11

\newif\if@fewtab\@fewtabtrue
{\count255=\time\divide\count255 by 60
\xdef\hourmin{\number\count255}
\multiply\count255 by-60\advance\count255 by\time
\xdef\hourmin{\hourmin:\ifnum\count255<10 0\fi\the\count255}}
\def\ps@draft{\let\@mkboth\@gobbletwo
    \def\@oddfoot{\hbox to 7 cm{\tiny \versionno
       \hfil}\hskip -7cm\hfil\rm\thepage \hfil {\tiny\draftdate}}
    \def\@oddhead{}
    \def\@evenhead{}\let\@evenfoot\@oddfoot}
\def\draftdate{\number\month/\number\day/\number\year\ \ \ \hourmin }

\global\def\draftcontrol{0}

\def\draftcite#1{\ifnum\draftcontrol=1#1\else{}\fi}
\def\@lbibitem[#1]#2{\item{}\hskip -3\hbox to 2cm
{\hfil$\scriptstyle\draftcite{#2}$}\hskip
1cm[\@biblabel{#1}]\if@filesw
     {\def\protect##1{\string ##1\space}\immediate
      \write\@auxout{\string\bibcite{#2}{#1}}}\fi\ignorespaces}

\def\@bibitem#1{\item\hskip -3cm \hbox to 2cm
{\hfil {\footnotesize\draftcite{#1}}}\hskip 1cm
\if@filesw \immediate\write\@auxout
       {\string\bibcite{#1}{\the\value{\@listctr}}}\fi\ignorespaces}

\def\citen#1{\if@filesw \immediate\write \@auxout {\string\citation{#1}}\fi%
\@tempcntb\m@ne \let\@h@ld\relax \def\@citea{}%
\@for \@citeb:=#1\do {\@ifundefined {b@\@citeb}%
    {\@h@ld\@citea\@tempcntb\m@ne{\bf ?}%
    \@warning {Citation `\@citeb ' on page \thepage \space undefined}}%
    {\@tempcnta\@tempcntb \advance\@tempcnta\@ne
    \setbox\z@\hbox\bgroup\ifcat0\csname b@\@citeb \endcsname \relax
    \egroup \@tempcntb\number\csname b@\@citeb \endcsname \relax
    \else \egroup \@tempcntb\m@ne \fi \ifnum\@tempcnta=\@tempcntb
    \ifx\@h@ld\relax \edef \@h@ld{\@citea\csname b@\@citeb\endcsname}%
    \else \edef\@h@ld{\hbox{--}\penalty\@highpenalty
    \csname b@\@citeb\endcsname}\fi
    \else \@h@ld\@citea\csname b@\@citeb \endcsname \let\@h@ld\relax \fi}%
\def\@citea{,\penalty\@highpenalty\hskip.13em plus.13em minus.13em}}\@h@ld}
\def\@citex[#1]#2{\@cite{\citen{#2}}{#1}}%
\def\@cite#1#2{\leavevmode\unskip\ifnum\lastpenalty=\z@\penalty\@highpenalty\fi%
  \ [{\multiply\@highpenalty 3 #1%
  \if@tempswa,\penalty\@highpenalty\ #2\fi}]}   %
\makeatother % end of CITE.STY

\catcode`\@=12

\def\Ad            {{\rm Ad}}
\def\be            {\begin{equation}}

\def\dl            {\mathbb }
\def\ee            {\end{equation}}

\newcommand\erf[1] {(\ref{#1})}

\def\futnote#1     {\footnote{~#1}\ }
\def\G             {{\rm G}}
\def\g             {{\bf g}}

\long\def\labl#1   {\label{#1}\ee \ifnum\draftcontrol=1
                   \mbox{ }\\[-12 mm]\query{#1}\\[5 mm] \fi}
\def\lie           {Lie algebra}

\long\def\query#1{\hskip 0pt{\vadjust{\everypar={}\small\vtop to 0pt{\hbox{}%
     \vskip -13pt\rlap{\hbox to 49.0pc{\hfil{\vtop{\hsize=8pc\tolerance=6000%
     \hfuzz=.5pc\rightskip=0pt plus 3em\noindent#1}}}}\vss}}}}%
\def\reals          {{\mathbb R}}
\def\rmd            {{\rm d}}
\newcommand\sect[1]{\section{#1}}
\def\T             {{\rm T}}
\def\Tr            {{\rm Tr}}

\def\zet           {{\dl Z}}

\documentclass[12pt]{article} \usepackage{amssymb,amsfonts,latexsym}
\usepackage[dvips]{graphics} \usepackage{epsfig}
\begin{document}
%%% for DRAFT versions, suppress in definitive version
  %\draft
  %\version\versionno

%%%%%%%%%%%%%%%%%%%%%%%%%%%%%%%%%%%%%%%%%%%%%%%%%%%%%%%%%%%%%%%%%%%%%%%

\begin{flushright}  {~} \\[-1cm]
{\sf hep-th/0108201}\\{\sf CPTH-S035.0701} \\ {\sf PAR-LPTHE 01-36}
\\[1mm]
{\sf August 2001} \end{flushright}

\begin{center} \vskip 14mm
{\Large\bf FLUX STABILIZATION IN COMPACT GROUPS} \\[20mm]
{\large Pedro Bordalo$\;^1$ \,, Sylvain Ribault $\;^2$
\ and \ Christoph Schweigert$\;^1$}
\\[8mm]
$^1\;$ LPTHE, Universit{\'e} Paris VI~~~{}\\
4 place Jussieu\\ F\,--\,75\,252\, Paris\, Cedex 05 \\[5mm]
$^2\;$ Centre de Physique Th{\'e}orique ~~~~{}\\
Ecole Polytechnique \\ F\,--\,91\,128\, Palaiseau \\
\end{center}
\vskip 18mm
\begin{quote}{\bf Abstract}\\[1mm]
We consider the Born-Infeld action for symmetry-preserving, orientable 
D-branes in compact group manifolds. We find classical solutions that obey 
the flux quantization condition. They correspond to conformally invariant 
boundary conditions on the world sheet. We compute the spectrum of quadratic 
fluctuations and find agreement with the predictions of conformal 
field theory, up to a missing level-dependent truncation. Our results
extend to D-branes with the geometry of twined conjugacy classes; they
illustrate the mechanism of flux stabilization of D-branes.
\end{quote}
\newpage

%%%%%%%%%%%%%%%%%%%%%%%%%%%%%%%%%%%%%%%%%%%%%%%%%%%%%%%%%%%%%%%%%%%%%%%

\sect{Introduction}

The study of D-branes has deepened our understanding of both
string theory and conformal field theory. Indeed, much of the
usefulness of D-branes comes from the fact that they can be investigated from
two rather different point of views: on the one hand side they are
described by boundary conditions on open strings. This allows to
investigate them using two-dimensional conformal field theories
for which many exact methods are available. On the other hand, their
worldvolume theories can be studied with conventional field theoretical
methods; this gives rise, in particular, to a rich interrelation with
Yang-Mills theories.

D-branes in group manifolds are a particularly tractable testing ground; 
they have recently received much attention. A large class of D-branes in 
WZW theories that is compatible with the symmetry encoded in the non-abelian 
currents has a rather simple description: their worldvolumes are (twined) 
conjugacy classes that obey an integrality condition. This result has
first been established by two-dimensional methods \cite{alsc2,fffs}.
It has been rederived from the Born-Infeld action,  
for ${\rm SU}(2)$ \cite{bads} and for ${\rm SU}(N)$ \cite{mms}. 

One surprisingly strong result of \cite{bads} was the good agreement
of the results obtained from the Born-Infeld action with the exact results
from conformal field theory. Indeed, up to a shift of the weights by the Weyl 
vector and of the level by the dual Coxeter number, the position
of the stable solutions of the Born-Infeld action coincide
with the exact CFT results.  Moreover, the spectrum of quadratic fluctuations
is close to the spectrum of boundary fields; the latter, however,
is truncated at finite level, an effect that is not reproduced by the 
Born-Infeld action. There have been speculations that the good agreement of the
results in the case of ${\rm SU}(2)$ is due to the fact that the supersymmetric
version of this theory appears in the description of the NS-fivebrane so
that effectively a hidden supersymmetry is responsible for the agreement.
In the present note, we use the Born-Infeld approach to investigate D-branes 
on general compact Lie groups and see that the
agreement holds in general. This shows in particular that the agreement
should not be seen as a miracle of string theory, but rather of
conformal field theory. (This can already be seen from the simple fact that
the Virasoro central charge of these theories can be 
arbitrarily high so that most of them cannot appear as building blocks
of string theories.)

It is known that compact connected Lie groups are quotients of the product 
of a semi-simple simply-connected
group and a torus by a subgroup of the center. The flux quantization
mechanism \cite{bads} requires a non-constant B-field; for simplicity
we therefore suppress the toroidal part in our discussion; 
our results, however, obviously extend to an arbitrary real compact Lie group.
 An important aspect of our
considerations is that we only use those general group theoretic methods that 
can also shed light on more general, non-compact groups.

This note is organized as follows: in Section 2 we briefly discuss the
Born-Infeld action and derive a useful description of the 
Kalb-Ramond background field $B$. In Section 3 we impose
the flux quantization condition to find candidates for solutions of the 
equations of motion for the Born-Infeld action. 
We compute first order and second order fluctuations around
these configurations. The first order fluctuations vanish which shows
that we have indeed found solutions. The spectrum of the second order
fluctuations is positive definite, which shows the stability of the solutions.
We compare the spectrum of quadratic fluctuations to the spectrum of operators 
on the worldvolume of the D-brane that has been obtained in conformal field 
theory calculations. In Section 4 we briefly sketch the generalization
of our results to D-branes that are twined conjugacy classes. We finally
present our conclusions. Some calculations are relegated to an appendix.

\bigskip

After completion of this work, related results for $\G={\rm SU}(N)$
have appeared in appendix A of \cite{mms}. In particular, similar expressions 
for the background fields $B$ and $F$ have been presented; our results include
also expressions for the background fields that are adapted to the investigation
of twisted boundary conditions.

\section{The Born-Infeld action and the background fields}

The Born-Infeld action is a generalization of the minimal surface problem
to spaces $M$ that carry not only a metric $G$ but also a two-form
gauge field $B$ with three-form field strength $H=\rmd B$. More precisely,
one considers embeddings $p$ of a space $\Sigma$ with a two-form field strength
$F=\rmd A$ into $M$, with the action
\be S_{BI}(p, F) = \int_\Sigma \sqrt{\det(p^* G + p^* B + 2\pi F)} \, . 
\labl{boinf}
The Born-Infeld action involves an integration over the world volume of the 
brane; we will therefore restrict ourselves to branes that are orientable.
\futnote{The known D-branes in WZW theories are not all orientable;
unorientable D-branes appear in non-simply connected groups. They are
quotients of conjugacy classes of the simply connected cover $\tilde\G$
that are invariant under the action of a subgroup of the center of $\tilde\G$.}
The action leads to a variational problem in which both the abelian gauge 
field $A$ and the embedding $p$ are varied. The fact that 
the two-form $F$ comes from a field strength implies the following 
quantization condition:
\be [F] \in H^2(\Sigma, \zet) \, , \labl{quant}
i.e. the class of $F$ must be an integral element of the second cohomology.
Actually, the field strength $F$ transforms under gauge transformations
$B \to B + \Lambda$ of the two-form gauge-field $B$ like
$F  \to F - \frac{1}{2\pi} p^* \Lambda$. Equivalently, the transformation
of its vector potential $A$ is  $A \to A + d\theta - \frac{1}{2\pi} 
p^* \alpha $, with $\theta$ a function and $\alpha$ a one-form 
such that $\rmd\alpha = \Lambda$. However,
every regular conjugacy class is contained in a single coordinate patch
for the $B$-field, which allows us to neglect large gauge transformations
for our purposes. For a more careful discussion of these issues we refer to 
\cite{fist8}.
We study the action \erf{boinf} in the case when $M = {\rm G}$ is a compact, 
semi-simple, connected, real Lie group and when $\Sigma$ is a 
homogeneous space of the form ${\rm G}/{\rm T}$, where ${\rm T}$ is a maximal 
torus of ${\rm G}$. 

\medskip

Our first task is to find explicit expressions for the background fields on 
the group manifold ${\rm G}$. To this end we need to fix a non-degenerate,
ad-invariant, bilinear form $\left\langle\cdot,\cdot\right\rangle$ on 
$\g={\rm Lie}\, \G$. 
Such a form is indeed a crucial ingredient to have a conformal field
theory with a group manifold as a target space;
we therefore pause to present a few comments
about it.  It has been shown \cite{fist} that a 
general, not necessarily compact, Lie group gives rise to a conformal 
field theory with all non-abelian currents as symmetries if and only if its 
Lie algebra admits such a form. (Note, though, that this form can be different 
from the Killing form; this is necessarily the case for non semi-simple
Lie groups.)

The existence of such a form also allows to reconcile the two seemingly
different sets of labels for boundary conditions. From conformal field
theory, it is known that conformally invariant boundary conditions 
correspond quite generally to representations of the chiral symmetry 
(see \cite{fuSc16} for the most general statements). In our case, the D-branes 
on a group manifold correspond to (twisted) representations of the group. 
The space-time approach, on the other hand, gives conjugacy classes which,
of course, are the orbits of the
{\em adjoint} action of a Lie group ${\rm G}$ on itself. Representations, 
in contrast, can be obtained from the quantization of
{\em co-adjoint} orbits (for a review see \cite{Kiri}). Understanding
this relation thus requires a natural relation between
the adjoint and the co-adjoint action of ${\rm G}$ on its \lie\ $\g$.
This is achieved by a non-degenerate, invariant, bilinear
form $\left\langle \cdot,\cdot\right\rangle$ on $\g$. 

\medskip

The Lie algebra $\g$ is a direct sum of pairwise orthogonal simple ideals,
$\g= \oplus_{i=1}^n \g_i$. On each simple ideal, we normalize the Killing form
such that for $x,y$ in the same simple ideal $\g_i$ we have
\be \kappa_i( x , y ) = - \frac{1}{8\pi g_i^{\vee}} \Tr(ad_x ad_y) \,,
\labl{bracket}
where $g_i^{\vee}$ is the dual Coxeter number of $\g_i$. A general 
invariant bilinear form on $\g$ is of the form
\be \left\langle v,w\right\rangle = \sum_{i=1}^n k_i \kappa_i(P_i v, P_iw)
\,, \labl{metr}
where $P_i$ is the orthogonal projection to the ideal $\g_i$. As is well-known,
consistency of the WZW action requires the numbers $k_i$ to be quantized. 
For a simply connected group, they have to be integers. For non-simply 
connected groups, further conditions have to be imposed: e.g.\ for 
$\G={\rm SO}(3)$, $k$ has to be even. The $n$-tuple of non-negative integers 
$(k_i)$ will be called the level.

Let us now give the background fields in terms of the bilinear form
\erf{metr}. We endow the manifold ${\rm G}$ with the ${\rm G}$-bi-invariant 
metric
\be G(u,v) =  \left\langle g^{-1} u , g^{-1} v \right\rangle ,  \labl{metric}
and the three-form field strength 
\be H(u,v,w) = - \left\langle g^{-1} u, [g^{-1} v,g^{-1} w] \right\rangle  \, , 
\labl{H-field} 
where $u,v,w \in T_g {\rm G}$. Due to the normalization of the bilinear
form \erf{metr}, the three-form $H$ is in $H^3(\G,\zet)$.

We wish to find a two-form potential $B$ for $H$. To this end we use the
well-known fact that for every compact connected Lie group the map
\be \begin{array}{ll}
q\,: \quad & {\rm G}/{\rm T} \times {\rm T} \to {\rm G}\\[.2em]
           & q(gT, t) := gtg^{-1} 
\end{array}\ee
is of mapping degree $|W|$, where $|W|$ is the number of
elements in the Weyl group $W$ of ${\rm G}$. In particular, $q$ is surjective.

Recall that a regular element of ${\rm G}$ is an element that is 
contained in only one maximal torus. In this note, we will restrict
ourselves to D-branes whose world-volume is contained in the subset
${\rm G}_r$ of regular elements. (The set regular elements ${\rm G}_r$ forms 
an open dense subset of ${\rm G}$; the codimension of its complement is
at least 3.)
The elements of a fixed conjugacy class are either all regular or not; in the 
former case, the conjugacy class will be called regular. Consider
${\rm T}_r = {\rm G}_r\cap {\rm T}$, i.e.\ the regular elements in the maximal 
torus; these elements constitute the interior of the Weyl chambers, on which the
action of the Weyl group $W$ is free. Each regular conjugacy class intersects 
the maximal torus in ${\rm T_r}$; the different intersections form an orbit
of the Weyl group. The map $q$ thus allows us to see $\G_r$ topologically 
as a product of $\G/\T$ over the interior of one Weyl chamber,
i.e.\ over $\T_r / W$. The notation
$$ q_t( gT) := q(gT, t) $$
for a family of maps from ${\rm G}/{\rm T}$ to ${\rm G}$ will come in handy.

To write down $B$, we introduce two more objects. The first object
is a family $F_t$ of two-forms on ${\rm G}/{\rm T}$, parametrized by an element
$t\in \T_r$:
\be 2\pi F_t(u,v) : = 2 \left\langle \exp^{-1} t , [h^{-1}u, h^{-1} v] 
\right\rangle \, , \labl F
where $h\in{\rm G}$ is an arbitrary representative of the point 
$hT \in \G / \T$ and $u,v \in T_{hT} (\G / \T)$. 
For each value of the parameter $t$ the two-form $F_t$ on $\G/\T$ is closed.
The vector $\exp^{-1} t\in\g$ is, of course, only defined up to elements of the 
lattice ${\rm Ker}(\exp)$. (The ambiguity in \erf{F} in choosing a preimage for 
the exponential map will be discussed at the end of this section.) 
For an arbitrary compact group
this lattice is called the integral 
lattice. For simply-connected groups
it coincides with the co-root lattice, whereas for non simply-connected groups
the integral lattice contains the co-root lattice. The finest possible integral
lattice is the one of the adjoint group and equals the 
co-weight lattice. 

The second object we need is a two-form $\omega$ on ${\rm G}_r$ \cite{alsc2}. 
We define $\omega_g$ to be
non-zero only on vectors $v\in T_g {\rm G}$ for which $g^{-1} v$
is in the image of $1-\Ad_g$, i.e. on the vectors tangent to the
conjugacy class that contains $g$. For such vectors, we set
\be \omega_g(u,v) = \left\langle g^{-1} u, \frac{1+\Ad_g}{1-\Ad_g} 
g^{-1} v \right\rangle . \ee

We then give the two-form potential $B$ on $\G_r$ by its pull-back to
$\G/\T \times \T_r$:
\be q^* B_{|(g\T,t)} = q^* \omega_{|(g\T,t)} - 2\pi (F_t)_{|gT} \, . \labl B
where the two-form $F_t$ on $\G/\T$ is extended to a two-form on
$\G/\T \times \T_r$ by setting it to zero on directions tangent to
$\T_r$. One can check that indeed $\rmd B = H$ on ${\rm G}_r$
(for details, see the appendix).

To conclude this section, we explain why choosing another Lie algebra
element for $\exp^{-1} t$ in the definition of $F_t$ \erf{F} does not affect 
the physics of the system. We already noted that the fields $B$ and $F$ 
are subject to large gauge transformations that have, of course, to be
compatible with the quantization of the level which guarantees that
$[H]\in H^3(\G,\zet )$.
One can check that a change of the Lie algebra element $\exp^{-1} t$
in \erf F corresponds to such an allowed gauge transformation, if and
only if this change is in the integral lattice. Changing the 
representatives for $\exp^{-1} t$ thus amounts to performing a large
gauge transformation.
\futnote{Note that our gauge choice respects the quantization of  
$[H]$, in the sense defined in \cite{bads}; our $B$-field vanishes at
the origin. }

\section{Flux quantization and classical solutions}

To obtain candidates for a solution of the equations of motion of \erf{boinf}
whose worldvolume is of the form $\G/\T$, 
we use $q_t$ as the embedding and $F_t$ as the field strength. This
idea, as well as the form of $F_t$ are inspired by Kirillov's method
of coadjoint orbits \cite{Kiri}. To qualify as a field strength, $F_t$ has 
to be quantized as in \erf{quant}. We first investigate for which $t$ this is 
the case.

The condition \erf{quant} amounts to the statement that the integral
of the two-form $F_t$ over any two-sphere in the world volume ${\rm G}/{\rm T}$ 
of the D-brane is an integer. A basis for $H_2(\G/\T,\zet)$ can be obtained
as follows: for any simple root $\alpha$, consider the corresponding
${\rm su}(2)$ subalgebra $\g_\alpha$ spanned by the co-root $H^\alpha$
and suitable linear combinations of the step operators $E^{\pm\alpha}$.
Its image in ${\rm G}$ under the exponential map is either isomorphic
to ${\rm SU}(2)$ or to ${\rm SO}(3)$. In both cases, quotienting by
${\rm T}$ gives a two-sphere $S_\alpha$ in ${\rm G/T}$. The spheres 
$S_\alpha$ generate $H_2(\G/\T)$. This reduces the problem to the well-known
case of $SU(2)$: the integral
$$ \int_{S_\alpha} F_t $$
is an integer iff $\exp^{-1}t$ is in the lattice dual to the co-weight lattice;
here the duality is with respect to the form \erf{metr}.
If we use the standard Killing form to 
identify the Cartan subalgebra $\g_0$ of $\g$ with its dual $\g_0^*$,
the weight space, then the component of $P_i (\exp^{-1} t)$ in the ideal 
$\g_i$ is of the form $2\pi\lambda/k_i$, where $\lambda$ is an integral
weight of $\g_i$. 

To see that we have indeed solutions of the classical equations of motion,
we compute the first order and second order fluctuations of the action under
variations of the gauge field $A$ on the brane and of the embedding.

We begin with first order fluctuations: with the notation
$M_t=q_t ^* G + q_t ^* B +2\pi F_t$, we have 
\be
\delta^{(1)} S_{BI} (q_t,F_t) = \frac{1}{2} \int _{{\rm G}/ {\rm T}} 
\sqrt{\det \left( M_t \right)} 
 \Tr \left( M_t^{-1} \delta M_t\right). 
\labl{var}
For a first order variation $\delta A$ of the one-form gauge field, one has
\begin{eqnarray}
\delta^{(1)}_F S_{BI} (q_t,F_t) &  =  & \int _{\G / \T} \sqrt{\det 
\left( M_t \right
)} \Tr \left( M_t ^{-1} \rmd (\delta \! A) \right) \nonumber \\
    &  =  & - \int _{\G / \T} \Tr \left( \delta \! A \ \rmd \left(
        \sqrt{\det \left( M _t \right)} M_t ^{-1} \right) \right).
    \nonumber
    \label{deltaF}
\end{eqnarray}
An explicit calculation (cf.\ the appendix) shows that
$$  \rmd \left( \sqrt{\det \left( M_t \right)}M_t ^{-1}
 \right)=0 \,, $$
so that first order fluctuations of the gauge field around our configurations
vanish. As for the fluctuations of the embedding $p$, one can show that
for any fluctuation $\delta_p M_t$ one has the identity
$$ \Tr \left( M_t ^{-1} \delta_p M_t \right) = 0  \,, $$
which implies that $\delta_p^{(1)} S_{BI} =0$.
This shows that we have indeed found classical solutions. Their positions 
form a lattice which we write in terms of the weight lattice $L^w_i$ of each 
simple ideal $\g_i$, up to the above-mentioned identification of $\g_0 $ with
$\g_0^*$:
\be L=\oplus_{i=1}^n \frac{2\pi}{k_i}L^w_i. \labl{lattice}
Recall that the set of conjugacy classes is parametrized by
$\T_r / W$, or, via the exponential map, by $\g_0 / \hat{W}$, where
$\hat W$ is the affine Weyl group, $\hat W = W \ltimes {\rm Ker}(\exp)$. 
Therefore, elements of $L$ that are related by the action of the affine
Weyl group are in fact identical solutions. Inequivalent solutions are
thus parametrized by the coset $L/\hat{W}$. 

Our result for the positions coincides,
up to a shift of the weights by the Weyl vector and a shift of 
the level by the dual Coxeter number, with the exact results \cite{fffs}
of conformal field theory.

\bigskip

To compute the quadratic fluctuations, we introduce coordinates $x^i$ on
$\G/\T$ and a lift $h$ from $\G/\T$ to $\G$, i.e.\ $h=h(x^i)\in\G$, and 
coordinates $\psi^\alpha$ on $\T$, i.e.\ $t(\psi^\alpha)\in\T$. For any choice 
of $t\in\T$, we then define $e_i$ 
as the orthogonal projection of 
\be m_i(x):= (\partial_{x^i} h) h^{-1} \labl{mdef}
on ${\rm Im}(1-\Ad_{hth^{-1}})$. Neither the vector $m_i$ nor the image depend
on $t$, since $t$ is a regular element of the maximal torus. 
We now use the invariant metric \erf{metr} to define the metric tensor
\be \gamma_{ij} (x) = \left\langle e_i(x) ,e_j(x)\right\rangle \labl{gamma}
on the tangent space of $T_{h(x)\T}(G/T)$. Moreover, considering 
$ g = h t(\psi^\alpha) h^{-1}$ as a function of $\psi^\alpha$ gives us
vectors 
$$ u_\alpha := g^{-1} \partial_{\psi^\alpha} g \,, $$
that are a basis of ${\rm Ker}(1-\Ad_g)$, the orthogonal complement of
${\rm Im}(1-\Ad_g)$ in $\g$. The basis 
$(e_i)$ of ${\rm Im}(1-\Ad_g)$ allows us to consider any two-tensor
$A_{ij}$ on $\G/\T$ as a linear operator $A$ on ${\rm Im}(1-\Ad_g)$, such 
that $$A_{ij}=\left\langle e_i ,A(e_j) \right\rangle ;$$ then its trace is 
$\Tr (A)=\left\langle e^i ,A(e_i)\right\rangle $ (indices are raised by $\gamma_{ij}$) 
and the inverse tensor corresponds to the inverse operator. 

Considering the $\psi^\alpha$ as real functions on $\G/\T$
describes a general embedding of $\G/\T$ in the neighborhood of a conjugacy
class.  We can thus parametrize an infinitesimal variation of the embedding
$p$ around the fixed embedding $q_t$ by functions $\delta\psi^\alpha$.
The fluctuations of the gauge field are described in terms of fluctuations
$\delta A$ of the gauge potential $A$. Note that the functions
$\delta\psi^\alpha$ are scalar functions on $\G/\T$ while the function
$\delta A$ is a one-form on $\G/\T$.

To write down the equations of motion, we introduce a covariant derivative
$\nabla$ on $\G/\T$ which in the coordinates $x^i$ reads
\be \nabla^i = \frac1{\sqrt{\det\gamma}} \partial_j \sqrt{\det\gamma}
\gamma^{ij} \, . \ee

The group $\G$ acts on the space of functions on $\G/\T$ by left translation
and thus provides a quadratic Casimir operator $\Box$ which coincides with
the Laplacian of the connection $\nabla^i$:
$$ \Box = \nabla^i \partial_i \, . $$

We are now ready to give the second order fluctuations, 
\begin{eqnarray}
\delta^{(2)} S_{BI} (t) &  = &  \int _{\G / \T} dx^i \sqrt{\det \left(
     M_t \right)}  \times  \nonumber \\
 & \times &  \left[ \frac{1}{8}   \gamma^{ij} 
 \partial_i \delta\!\psi^{\alpha} \partial_j
\delta\!\psi^{\beta} \delta_{\alpha \beta} - \frac{1}{8}\Tr
                 \left( ad_{u_{\alpha}} ad_{u_{\beta}} \right)
\delta\!\psi^{\alpha} \delta\!\psi^{\beta} \right. \label{d2s}\\
 & - &\left.  \frac{1}{4} \Tr \left( \delta\! F \frac{2}{1-\Ad_g} \delta\! F
\frac{2}{1-\Ad_g}\right) - \frac{1}{8} \Tr \left(\delta\! F
            ad_{u_{\alpha}}\right) \delta\! \psi^{\alpha}
 + \frac{1}{8} \Tr ^2 \left(\frac{2}{1-\Ad_g} \delta\! F  \right)
     \right].  \nonumber
\end{eqnarray}
This implies the equations of motion:
\begin{eqnarray}
\delta\!\psi^{\alpha}: & \ \ &  \delta _{\alpha \beta} \Box
\delta\!\psi^{\alpha}  +  \Tr\left( ad_{u_{\alpha}} ad_{u_{\beta}}
\right) \delta\!\psi^{\alpha} +
\frac{1}{2} \Tr \left( \delta\!F ad_{u_{\beta}}  \right) = 0 \, ,
\label{eq.1} \\
\delta\!A_j: & \ \ &  -4(ad_{u_{\alpha}})^{ij} \partial_j
\delta\!\psi^{\alpha} + \frac12\left\langle e^i,[e^k,e^l]\right\rangle 
\delta\!F_{kl} + \nabla^k
\delta\!F _k\,^i = 0 \, .  \label{eq.2}
\end{eqnarray}
These equations should be complemented by the gauge condition
\be \nabla (\delta A) = 0 \, ,   \labl{gauge1}
which is invariant under the action of $\G$. It should be appreciated
that neither the equations of motion nor the gauge condition depend
on the position $t$ of the brane. This is in accordance with the result
from conformal field theory \cite{fffs} that in the large level limit
the fields on a brane do not depend on its position.

\medskip

Using the fact that $e_i, u ^{\alpha}$ form a basis of $\g$, we parametrize 
the fluctuations $\delta A$ and $\delta\psi^\alpha$ in terms of a single 
function $f:\G / \T \rightarrow \g$:
\be
\delta\!A_i = 2 \left\langle e_i, f  \right\rangle  \ , \quad  \delta\!\psi^{\alpha}
= \left\langle u ^{\alpha}, f \right\rangle,  \labl{ansatz}
The gauge condition (\ref{gauge1}) then translates into
\be \left\langle e_i , \partial^i f \right\rangle = 0  \labl{gauge2} 
while the equations of motion (\ref{eq.1},\ref{eq.2}) become
\be
\left\langle  u_{\beta},  \Box f \right\rangle = 0 \,, \quad  
\left\langle e_i, \Box f  \right\rangle = 0  \ee
which amounts to
\be \Box f = 0 \, .  \labl{eq.f1}

We now consider how the solutions of \erf{eq.f1} and \erf{gauge2} transform 
under the left action of $\G$. The function $f$ itself transforms in the
adjoint representation $R_\theta$. Since the differential operator in 
\erf{eq.f1} is a singlet under the action of $\G$, 
the solutions of \erf{eq.f1} come in 
representations of the form $R_\theta\, \otimes \, R_0$ where $R_0$ is any 
representation of $\g$ with vanishing second order Casimir eigenvalue.
The gauge condition \erf{gauge2}, on the other hand, is a $\g$-covariant map 
from $R_\theta\, \otimes \, R_0$ to $R_0$; thus it removes one 
representation isomorphic to $R_0$ from the solutions $R_\theta\otimes R_0$.

This result is, in itself, of limited interest as long as we take $\G$
to be a compact group: in this case, the only representation of
Casimir-eigenvalue 0 is the trivial, one-dimensional, representation.
However, all our computations continue to hold if we take the group 
to be the product $\G_{{\rm tot}} = \G\times \reals$ 
of a compact group $\G$ with a time-like factor $\reals$ with  negative metric. 
(One could also take a product with some other non-compact group, see
\cite{bape,peri}.)
The D-brane is supposed to extend along the time-like 
direction, while its spatial position is still parametrized by the
regular elements of the maximal torus of the compact group $\G$. The 
contribution of $\reals$ to the total Casimir eigenvalue of 
$\G_{{\rm tot}} = \G \times\reals$ takes arbitrary negative values 
that can cancel any positive contribution from $\G$. In this case, all
unitary representations $R$ of $\G$ appear in the spectrum. 

The fluctuation
spectrum including a timelike direction can be compared with results from 
conformal field theory, as in \cite{bads}.
We read off that the multiplicity of $R_\theta\, \otimes \, R \, - \, R$ in the 
spectrum of quadratic fluctuations equals the multiplicity of $R$ in the 
space of functions on $\G_{{\rm tot}}/\T$, which in turn is equal to the 
multiplicity of the corresponding representation in the space of functions
on $\G/\T$. This result differs, as in the case of ${\rm SU}(2)$, from the 
exact CFT result \cite{fffs}: for the latter, a level-dependent truncation 
of the representations $R$ is found. Only representations that are 
integrable at the given level appear in the exact result.
The energy eigenvalues of the quadratic fluctuations are of the form $C_R/k$, 
where $C_R$ is the
eigenvalue of the second order Casimir operator on the representation
$R$. They are positive, which shows the stability of our solutions; moreover,
they agree, as in the case of ${\rm SU}(2)$, with the conformal weights in
the limit of large $k$.

\medskip

To discuss the energies of the branes themselves, let us restrict for 
simplicity to a single simple factor.
The energy $E_\lambda$ of a D-brane at $2\pi\lambda/k$ is proportional
to the quantum dimension of the corresponding boundary condition:
\be D_{\lambda} = \prod_{\alpha > 0 } 
\frac{\sin \left(  \frac{\pi(\lambda + \rho, \alpha )}{k + g^{\vee}} 
\right)
}{\sin \left(  \frac{\pi(\rho, \alpha )}{k + g^{\vee}} \right)}   \, ,
\labl{qdim}
where the product is over the positive roots 
$\alpha$, and the parenthesis
denote the standard bilinear form on the Cartan subalgebra of $\g_i$, 
normalized such that the highest root has length squared 2.
Indeed, the direct calculation of $E_\lambda $ with 
$\exp^{-1} t=2\pi\lambda/k$ gives
\be E_\lambda =S_{BI}(q_{t},F_{t})\propto 
\int _{\G/\T}\sqrt{\det\gamma_{ij}}\sqrt{
\det (1-\Ad_{t})}, \ee 
where the operator $(1-\Ad_{t})$ has to be restricted to its image. 
We find 
\footnote{The $\lambda$-independent prefactors were computed in 
\cite{mms} for $\G={\rm SU}(N)$ and precise agreement was found in the 
limit of large level.}
\be E_\lambda\propto \prod_{\alpha > 0 }  \sin \left(\frac{\pi}{k} 
(\lambda,\alpha)\right) \, .  \ee
The results of this section can be summarized by the statement that,
in the limit of large level, the Born-Infeld results 
agree with the exact CFT results for all compact Lie groups.

\section{Generalization to twisted boundary conditions}

Our considerations can be easily generalized to D-branes that are twined
conjugacy classes \cite{fffs}  
${\cal C}^\omega(h) = \{ gh \omega^{-1}(g)\}$ where 
$\omega$ is an automorphism of $\G$. This automorphism relates left movers and 
right movers at the boundary and determines the automorphism type of the
boundary conditions. Twined conjugacy classes can be parametrized as
follows \cite{fffs}: any automorphism $\omega$ leaves at least one maximal torus
$\T$ of $\G$ invariant. Denote by $\T^\omega_0$ the connected component
of the identity of the subgroup of $\T$ that is left pointwise invariant
by $\omega$: for compact $\G$, this is a proper subgroup in the case of an 
outer automorphism. Up to the action of the subgroup of the Weyl group 
consisting of elements that commute
with $\omega$, $T^\omega_0$ parametrizes the twined conjugacy classes.
Regular twined conjugacy classes are diffeomorphic to $\G/\T^\omega_0$;
in particular, for $\omega$ an outer automorphism their dimension is greater 
than the dimension of regular conjugacy classes. 

The expressions for the background fields can be generalized as follows
(see also \cite{staN6}):
\be \omega_g (u,v) =  \left\langle g^{-1} u , \frac{1+\omega\circ \Ad_g}
{1-\omega \circ \Ad_g} g^{-1} v \right\rangle \, , \ee
where now $ g^{-1} u \in {\rm Im}(1-\omega\circ \Ad_g)$. The family
$F_t^\omega$ is now a family of two-forms on $\G/\T_0^\omega$ parametrized
by $\T^\omega_0$. Formally, we take the same expression as in \erf F, 
with the important difference that now $t$ is required to be in $\T_0^\omega$.

A similar argument as for the untwisted case shows that the quantization
condition \erf{quant} implies that $\exp^{-1} t$ is a fractional symmetric
weight, which agrees with the results in \cite{bifs,fffs}. (Note that
there is an additional shift of the weight lattice for ${\rm SU}(N)$
with $N$ odd which comes from extending the automorphism $\omega$ from
${\rm Lie} \, \G$ to the untwisted affine Lie algebra. It is not seen in
the Born-Infeld action, which does not have a direct relation to the
affine Lie algebra.)

We illustrate our results in the case of ${\rm SU}(3)$ at level $k=6$: 
figure \erf{fig} shows the intersections of the brane worldvolumes with 
the fundamental Weyl chamber in ${\rm Lie} \, \T$. Light circular vertices 
mark the intersections of singular branes with the boundary of the Weyl 
chamber. Full dark circles denote regular six-dimensional conjugacy classes 
on which a D-brane can wrap. Hexagonal marks give the intersection of the 
seven-dimensional worldvolume of twined conjugacy classes with the fundamental 
Weyl chamber; they are all contained in the one-dimensional subspace 
${\rm Lie} \, \T_0^\omega$. One and the same {\em twined} brane 
can intersect the fundamental Weyl chamber in more than one point; 
the corresponding hexagonal marks are linked by arrows.

\be \begin{picture}(200,176)(-5,0)
\scalebox{.31}{\includegraphics{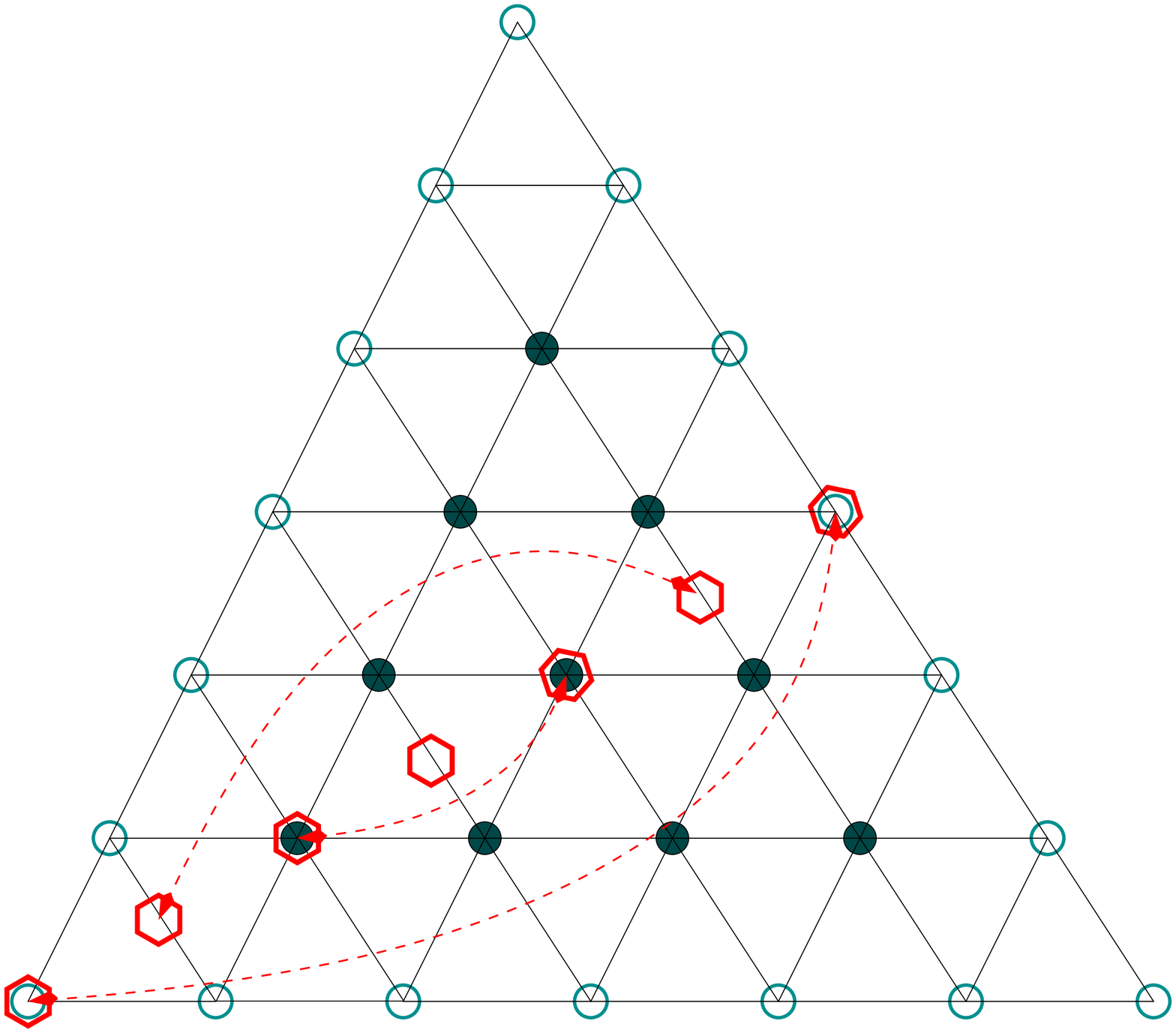}} \end{picture}
\labl{fig}

The calculation of the spectrum of quadratic fluctuations can
be generalized as well. To this end, it is necessary to realize that 
${\rm Lie}\T_0^\omega={\rm Ker}(1-\omega\circ \Ad_t)$. We then use the operator
$1-\omega\circ \Ad_t$ instead of $1-\Ad_t$ to define $e_i$s and $u_\alpha$s, as 
above. In every subsequent expression with a dependence in $\Ad_g$, the latter 
should be replaced by $\omega\circ \Ad_g$. In particular, we obtain a metric 
$\gamma_{ij}=\left\langle e_i,e_j \right\rangle $ on $\G/\T^\omega_0$ 
whose Laplacian coincides with the quadratic Casimir of the action of $\G$ on 
the space of functions on $\G/\T^\omega_0$  by left translation. The resulting 
spectrum is expressed in terms of eigenvectors of this operator to the 
eigenvalue zero.  The action of $\omega$ does not change the formal structure 
of the result, but, of course, modifies the multiplicities of the 
irreducible representations: we now have to decompose the space of functions on
$\G/\T^\omega_0$ (the 
multiplicities are given in (4.21) of \cite{fffs}).
The resulting multiplicities still match the corresponding annulus amplitude, 
up to the missing usual level-dependent truncation.

\section{Conclusions}

In this note, we have studied solutions of Born-Infeld theory on 
compact, connected Lie groups $\G$ that have the form of (twined) conjugacy
classes. We have found explicit expressions for the two-form gauge field
$B$ on regular group elements and the two-form field strength $F$ on
regular conjugacy classes. These expressions may be generalized to other cases,
including non-compact groups, provided their Lie algebra admits an invariant 
non-degenerate bilinear form. However, in the case of non-compact groups, even
if they are semi-simple, one has to be careful with the definitions of regular 
elements and Cartan subgroups.

The quantization condition on $F$ implies a quantization on the possible 
positions of the branes; they agree, up to a shift, with the exact CFT results.
The energies of the D-branes are proportional to 
the quantum dimensions of the corresponding boundary conditions.

Our results give us confidence that also on more general Lie groups (and
manifolds with possibly even less structure), solutions of the Born-Infeld 
action can provide information on boundary conditions of conformal field 
theories.

{\flushleft \bf Note added in proof:}
The metric (\ref{gamma}) on the conjugacy class is in fact the well-known 
open-string metric. As expected from CFT arguments \cite{bape}, it does 
not depend on the position of the D-brane. This phenomenon is called radius 
locking in \cite{ba}.

\vskip2em
\noindent
{\small
{\sc Acknowledgements}:\\ 
We are grateful to Nicolas Couchoud, Michel Duflo and Ingo Runkel for helpful
discussions and to Costas Bachas for a careful reading of the manuscript.}

\appendix

\section{Appendix}

From the main text, we recall the notations $h(x^i) \in \G$ for the local 
lift from $\G/\T$ to $\G$, $t(\psi^\alpha) \in \T$,  and finally
$g = hth^{-1} \in {\rm G}$.  We use latin indices for directions tangent to
$\G / \T$; they are raised and lowered with the metric 
$\gamma_{ij}$ (\ref{gamma});  
greek indices stand for directions tangent to the maximal torus.  We denote 
$C_g=C_t$ the conjugacy class of $g$ and $t$ respectively.

\medskip

{\flushleft \bf Calculation of $dB=H$:}

The two-form $B$ on ${\rm G}_r$ is defined by (\ref{B}); we rewrite
$F_t$ as  
\begin{eqnarray}
2\pi F_t = \left\langle exp^{-1} t, [h^{-1} \partial _i h, h^{-1} \partial _j h ]  
\right\rangle dx^i \wedge
dx^j \nonumber  \,  .
\end{eqnarray}
This two-form on ${\rm G}/{\rm T}$ is closed, since 
$\rmd([h^{-1}dh , h^{-1}dh])=-\rmd^2(h^{-1}dh)=0$. Next, we rewrite:
\begin{eqnarray}
\omega _g = \left\langle g^{-1}\partial _i g, \frac{1+Ad_g}{1-Ad_g}
  g^{-1}\partial _j g \right\rangle dx^i \wedge dx^j  \, .
\end{eqnarray}
It is known \cite{alsc2} that $\rmd\omega _{|_{C_g}}=H _{|_{C_g}}$; we thus
only need to calculate the components of the differential of 
($q_t ^* \omega -2\pi F_t$) tangent to the maximal torus: 
\begin{eqnarray}
\partial _{\alpha} 2\pi (F_t)_{ij}= 2\left\langle 
t^{-1} \partial _{\alpha} t, [h^{-1} \partial _i h,
h^{-1} \partial _j h ]   \right\rangle \,  ,  \\ 
\partial _{\alpha} \omega _{ij} = - \left\langle 
[t^{-1} \partial _{\alpha} t,\Ad_g \left( h^{-1} \partial
_{[i} h  \right)  ], h^{-1} \partial _{j]} h    \right\rangle \,   . 
\nonumber
\end{eqnarray}
They are to be compared to the expression for $H_{\alpha i j}$:
\begin{eqnarray}
H_{\alpha i j} = 2  \left\langle t^{-1} \partial _{\alpha} t, [h^{-1} \partial _i h ,
h^{-1} \partial _j h]  \right\rangle 
-  \left\langle t^{-1} \partial _{\alpha} t, [\Ad_g \left( h^{-1} \partial _i h \right),  h^{-1} \partial _j h] +
[h^{-1} \partial _i h , \Ad_g \left(  h^{-1} \partial _j h  \right)  ]     
\right\rangle \, .  \nonumber
\end{eqnarray}
The second term coincides with $\partial _{\alpha} \omega _{ij}$, while the
first one equals indeed $2 \pi \partial _{\alpha} (F_t) _{ij}$.
Finally, we notice that, for any index $M$, we have 
$H_{\alpha \beta M}=0$.  Since $B_{\alpha M}=0$, we have 
proved $\rmd B=H$.

\medskip

{\flushleft \bf Calculation of ${\rm Tr} \left( M_t ^{-1} \delta _{\alpha} M_t \right) 
= 0$ :}

We use the projection $e_i$ of
$\partial _i h \ h^{-1}$ on ${\rm Im}(1-\Ad_g)$ to write  
$(M_t)_{ij} = 2 \left\langle (1-\Ad_g)e_i, e_j \right\rangle $.
The geometrical fluctuation of $M_t$ is 
\begin{eqnarray}
\delta _{\alpha} (M_t)_{ij} = - 2  \left\langle [g^{-1} \partial _{\alpha} g,
(\Ad_g-1)e_i], e_j \right\rangle \,  .  \nonumber
\end{eqnarray}
Therefore, using the identity $(\Ad_g-1)g^{-1} \partial _{\alpha} g=0$ in the 
last equality, we obtain: 
\begin{eqnarray}
{\rm Tr} \left( M_t ^{-1}  \delta _{\alpha} M_t \right) & = &   
\left\langle \frac{1}{1-\Ad_g} e^j,e^i  \right\rangle  
 \left\langle (\Ad_g-1) e_i,[g^{-1} \partial _{\alpha} g, e_j]  \right\rangle \nonumber \\
& = &  \left\langle 
\frac{\Ad_g-1}{2 \Ad_g} g^{-1} \partial ^j g , [g^{-1} \partial _{\alpha} g,
\frac{1}{\Ad_g-1} g^{-1} \partial _j g]  \right\rangle  = 0 \nonumber  .
\end{eqnarray}

\medskip

{\flushleft
\bf Calculation of $\rmd \left(\sqrt{\det M_t} M_t ^{-1} \right)  = 0$ :}

We use the notation $T:= M_t ^{-1}$ to rewrite:
\be \rmd \left(\sqrt{\det M_t} M_t ^{-1} \right)=
\sqrt{\det \left( M _t \right)} \partial_j (M_t)_{rs} \left( T^{[s,r]}T^{[i,j]}-T^{[j,r]}T^{[s,i]} \right)\ dx_i.   \label{integrand}
\ee
In the sequel we will use $T^{ij}=\frac{1}{2}\left\langle g^{-1}\partial^ig,
(1-\Ad_g)g^{-1}\partial^jg\right\rangle $,
 where the indices are now raised by the metric $\left\langle g^{-1}\partial^ig,
g^{-1}\partial^j g \right\rangle $.
We also introduce 
the notation $ g^{-1} \partial _r \partial_j g =v_{rj} $. We can write
\begin{eqnarray}
\partial_j (M_t)_{rs}  = & & \frac{1}{2}  \left\langle g^{-1} \partial _j g,
[g^{-1} \partial_r g,\frac{1}{1-\Ad_g}g^{-1} \partial_s g] \right\rangle +  \nonumber \\
& & \left\langle g^{-1} \partial_r g, \frac{\Ad_g}{\Ad_g-1}
[g^{-1} \partial _j g,\frac{1}{1-\Ad_g}g^{-1} \partial_s g] \right\rangle  
+ \nonumber \\
& &\frac{1}{2} \left\langle g^{-1} \partial_r g, \frac{1}{\Ad_g-1} [g^{-1} 
      \partial_j g,g^{-1} \partial_s g] \right\rangle + \nonumber \\  
 & &  \left\langle v_{rj}, \frac{1}{\Ad_g-1}g^{-1} \partial_s g \right\rangle 
+ \left\langle g^{-1} \partial_r g, \frac{1}{\Ad_g-1}v_{sj} \right\rangle . 
\label{eq.dM} \end{eqnarray}
The contribution of the last line of \erf{eq.dM} to
\erf{integrand} vanishes by itself: using the fact that $v_{rj}=v_{jr}$, we
insert the first term of this line in \erf{integrand} and write the result
\begin{eqnarray}
& & \left\langle v_{rj}, \frac{1}{\Ad-g1}g^{-1} \partial_s g \right\rangle  
\left\langle g^{-1} \partial^s g, (1-\Ad_g) g^{-1} \partial^r g \right\rangle  \left\langle
g^{-1} \partial^j g,
(1-\Ad_g)g^{-1} \partial^i g \right\rangle  \nonumber \\
& = & -  \left\langle v_{rj}, g^{-1} \partial^r g \right\rangle   
\left\langle g^{-1} \partial ^j g, (1-\Ad_g) g^{-1} \partial ^i g \right\rangle \,  .  
\nonumber \end{eqnarray}
The second term of the same line 
gives a similar contribution, but with opposite sign.

The contribution of the first three lines of \erf{eq.dM} to \erf{integrand}
also vanishes. To see this, note that all the resulting terms are of the form
\begin{eqnarray}
 \left\langle  K(\Ad_g)g^{-1} \partial_i g , 
  [ L(\Ad_g)g^{-1} \partial^i g, M(\Ad_g) g^{-1} \partial^j g] \right\rangle , 
\nonumber \end{eqnarray}
where $K,L,M$ are some operators on $\g$ built of $\Ad_g$.  
Then $M(\Ad_g) g^{-1} \partial ^j g \in \mathrm{Im}(1-\Ad_g)$ whereas
$[ K(\Ad_g) g^{-1} \partial _i g,  L(\Ad_g) g^{-1} \partial ^i g ]  
\in \mathrm{Ker}(1-\Ad_g)$, 
and therefore all these terms vanish.   

%%%%%%%%%%%%%%%%%%%%%%%%%%%%%%%%%%%%%%%%%%%%%%%%%%%%%%%%%%%%%%%%%%%%%%%%
%\newpage
 \vskip2.6em
\small
 \newcommand\wb{\,\linebreak[0]} \def\wB {$\,$\wb}
 \newcommand\Bi[1]    {\bibitem{#1}}
 \newcommand\J[5]   {{\sl #5}, {#1} {#2} ({#3}) {#4} }
 \newcommand\PhD[2]   {{\sl #2}, Ph.D.\ thesis (#1)}
 \newcommand\Prep[2]  {{\sl #2}, preprint {#1}}
 \newcommand\BOOK[4]  {{\em #1\/} ({#2}, {#3} {#4})}
 \newcommand\iNBO[6]  {{\sl #6}, in:\ {\em #1}, {#2}\ ({#3}, {#4} {#5})}
 \def\adma  {Adv.\wb Math.}
 \def\aspm  {Adv.\wb Stu\-dies\wB in\wB Pure\wB Math.}
 \def\atmp  {Adv.\wb Theor.\wb Math.\wb Phys.}
 \def\bams  {Bull.\wb Amer.\wb Math.\wb Soc.}
 \def\comp  {Com\-mun.\wb Math.\wb Phys.}
 \def\ijmp  {Int.\wb J.\wb Mod.\wb Phys.\ A}
 \def\jgap  {J.\wb Geom.\wB and\wB Phys.}
 \def\jhep  {J.\wb High\wB Energy\wB Phys.}
 \def\joal  {J.\wB Al\-ge\-bra}
 \def\maan  {Math.\wb Annal.}
 \def\nupb  {Nucl.\wb Phys.\ B}
 \def\phlb  {Phys.\wb Lett.\ B}
 \def\phrd  {Phys.\wb Rev.\ D}
 \def\phrl  {Phys.\wb Rev.\wb Lett.}
 \def\pnas  {Proc.\wb Natl.\wb Acad.\wb Sci.\wb USA}
 \def\rvmp  {Rev.\wb Math.\wb Phys.}
 \def\AP     {{Academic Press}}
 \def\NH     {{North Holland Publishing Company}}
 \def\SV     {{Sprin\-ger Ver\-lag}}
 \def\WS     {{World Scientific}}
 \def\Ad     {{Amsterdam}}
 \def\Be     {{Berlin}}
 \def\NY     {{New York}}
 \def\Si     {{Singapore}}

\end{document}